# Optical absorption spectrum reveals gaseous chlorine in anti-resonant hollow core fibres

K. HARRINGTON,* R. MEARS, J. M. STONE, W. J. WADSWORTH, J. C. KNIGHT, AND T. A. BIRKS

*Centre for Photonics and Photonic Materials, Department of Physics, University of Bath, Bath, BA2 7AY, UK*
*kh302@bath.ac.uk*

**Abstract:** We have observed unexpected spectral attenuation of ultraviolet light in freshly drawn hollow core optical fibres. When the fibre ends are left open to atmosphere, this loss feature dissipates over time. The loss matches the absorption spectrum of gaseous (molecular) chlorine and, given enough time, the transmission spectrum of the fibre recovers to that expected from the morphological structure of the fibre. Our measurements indicate an initial chlorine concentration of 0.45 µmol/cm$^3$ in the hollow core, equivalent to 1.1 mol% $Cl_2$ at atmospheric pressure. © 2024 The Author(s)

## 1. Introduction

Due to the low overlap of guided light with the glass, anti-resonant hollow core fibres (HCFs) can guide at wavelengths where their solid-silica counterparts suffer from high attenuation or photo-darkening [1, 2]. Such fibres are widely considered to be comparatively unaffected by material limitations. Nevertheless, the hollow core is susceptible both to contamination and to changes over time, especially when left open to atmosphere. Known issues include water ingress [3], pressure changes [4], and contaminants such as hydrogen chloride (HCl) [2] and ammonium chloride [5]. The ro-vibrational absorption spectrum of HCl has long been observed in the transmission spectrum of hollow-core fibres in the mid-IR [2], and is attributed to the presence of chlorine originating from the low-OH F300 synthetic fused silica material from which such fibres are often made [6]. The advent of HCFs transmitting in the deep ultraviolet (UV), where gaseous chlorine has a strong absorption band, recently enabled us to demonstrate the presence of chlorine presumed to originate from the photo-dissociation of HCl in an HCF some months after its fabrication [7].

In this work we report the presence of molecular chlorine gas ($Cl_2$) in the core space of an HCF made from low-OH synthetic fused silica, immediately after its fabrication. An unexpected UV attenuation feature, matching the known absorption spectrum of molecular $Cl_2$, was observed to dissipate over a time consistent with diffusion from the ends of the fibre. Our findings therefore suggest that all hollow-core fibres drawn from low-OH silica may have optical properties that change with time. For fibres drawn to lengths of 100s of metres or more, this time scale would be many years. It may be possible to avoid this by fabricating the fibre from chlorine-free (typically high-OH) silica, although this may introduce other contaminants.

## 2. UV-guiding HCF fabrication and transmission

We and others [1,8,9,10] have previously reported hollow core fibres for the UV, with recent fibres guiding most of the ultraviolet range possible in air [8]. One example, made using the stack and draw method, is shown as an inset in Fig 1. It had an outer diameter of 80 µm, an inscribed core diameter of 13-15 µm, and capillaries with outer diameters of 7-8 µm and wall thicknesses of ~190 nm. We cut 2 metres of fibre (the fibre under test) from the middle of a 200 m length within a few hours of its fabrication. The ends were cleaved with a low cleave angle (<2 degrees) and left open to the atmosphere, with no further cleaving over the course of the experiment. Broadband light from a laser-driven light source (Energetiq EQ-99X) was coupled into one end of the fibre under test. The output spectrum was repeatedly measured every 5 minutes with a Bentham DTMc300 double monochromator and large-area UV photodetector, Fig 1. Although only a few of the traces were shown in this graph, all 840 traces were analysed and are available in the dataset [11].

The transmission spectrum of the fibre 4 days after the experiment began (shown as the top trace labelled '4 days' in Fig. 1) exhibited the familiar pattern of low-loss anti-resonant windows bounded by high-loss resonances centred at wavelengths given by

$$\lambda_m = \frac{2t}{m}\sqrt{n^2-1}$$

where resonance order $m$ is an integer, $t$ is the core wall (capillary) thickness and $n$ is the refractive index of silica [2]. Thus the high-loss bands at 220-240 nm and 380-460 nm were respectively the $m = 2$ and $m = 1$ resonances expected from the known wall thickness $t$.

However, the trace at the start of the experiment (marked 0 hours in Fig. 1) exhibited an additional high loss feature in the UV centred at ~325 nm and extending into the visible range, which cannot be explained by the structure of the fibre. The figure shows this feature steadily dissipating during the course of the experiment.

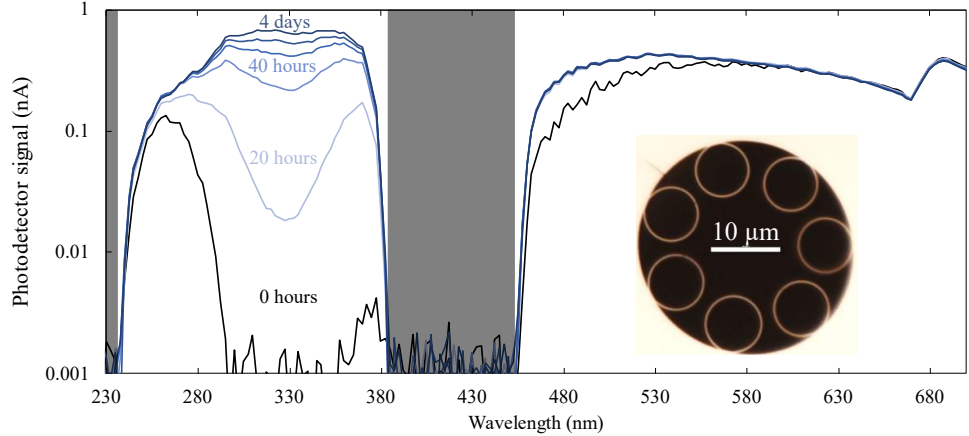

**Fig 1.** Transmission traces through 2 m of HCF over time, where the first 5 traces show snapshots of the dataset taken at 20 hour intervals. The final trace was taken after 4 days. The first trace (black) was taken shortly after the piece of fibre was removed from the 200 m length. The shaded regions of the wavelength ranges are the high loss resonant bands of the HCF, as expected from the structure. Inset: An optical micrograph of the cross-section of the fibre under test.

## 3. Chlorine concentration in the hollow core over time

The increase of transmission between an early trace (roughly 12 hours from cleaving the fibre, chosen because it is the first trace where the signal between 300-360 nm was above the noise floor), and the last trace taken after 4 days, is shown in Fig 2. Also shown is the reported absorption spectrum of gaseous $Cl_2$ [12]. The two curves match very closely, suggesting that absorption by $Cl_2$ is the cause of the loss. Similar results were obtained for traces recorded at other times. Other gases that might be present ($N_2$, $O_2$, $CO_2$, $H_2O$, $HCl$, $O_3$, $N_2O$, $NO$, $NO_2$) do not have such a feature, and whereas $N_2O_4$ does have a loss feature at 340 nm it also has a similarly-strong absorption at 260 nm that is entirely absent from our measurements [12].

The transmitted signal at the $Cl_2$ absorption peak at 330 nm is plotted in Fig 3 as a function of time, together with the corresponding mean concentration of $Cl_2$ in the core calculated using Beer-Lambert law. Repeating the calculation for 270 nm (where the signal is above noise even at the start of the experiment) to determine the change between 0 to 12 hours allows us to estimate that the initial mean concentration of $Cl_2$ in the hollow core was 0.45 µmol/cm$^3$, which is equivalent to 1.1 mol% at atmospheric pressure. (Since the initial pressure may be below atmospheric [4], the value in mol% is a lower bound.)

From the reported diffusion constant $D$ of $Cl_2$ in air [13] and the 1 m half-length $l$ of the fibre, the order-of-magnitude diffusion time $\tau \sim l^2/D$ should be a little less than a day [14]. This is consistent with the time scale in Fig. 3, implying that the $Cl_2$ dissipates mainly by diffusion from the fibre's ends. However, evidence of HCl in HCFs some time after their fabrication [2] suggests that some $Cl_2$ is also consumed by reaction with atmospheric water vapour travelling inwards by diffusion or pressure-driven flow. For a more robust understanding it may be possible to simultaneously observe the gas absorption lines of HCl in the mid-IR and of $Cl_2$ around 325 nm in an appropriately designed hollow optical fibre. In the present recently-fabricated fibre we did not observe the HCl-related generation of $Cl_2$ that we reported in older fibres [7], suggesting that HCl formation takes place over a longer time.

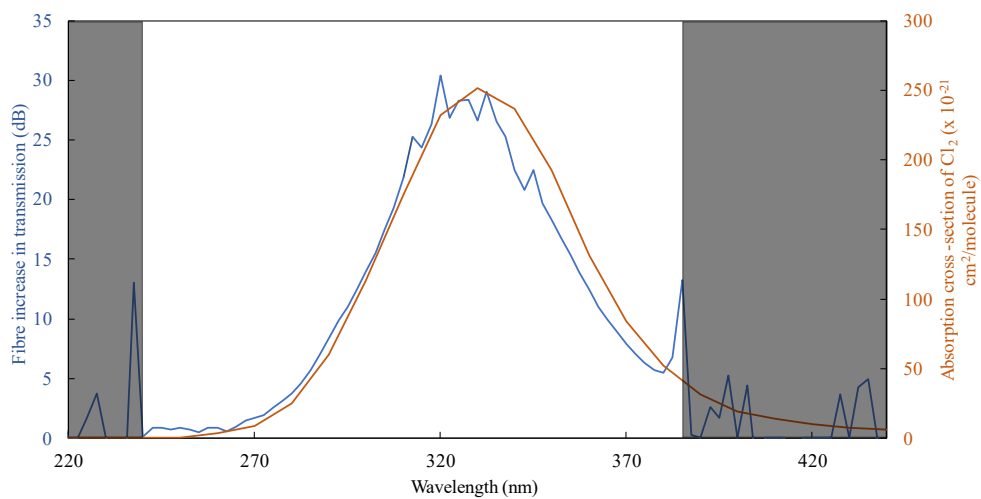

**Fig 2.** Increase in transmission through 2 m of fibre between 12 hours and 4 days from the start of the experiment, plotted with the absorption spectrum of gaseous chlorine [12]. The noisy shaded regions correspond to the high loss resonance bands of the HCF. The fibre transmission data may be spectrally shifted by ± 3 nm depending on alignment of the small fibre core to the slit in the monochromator set up.

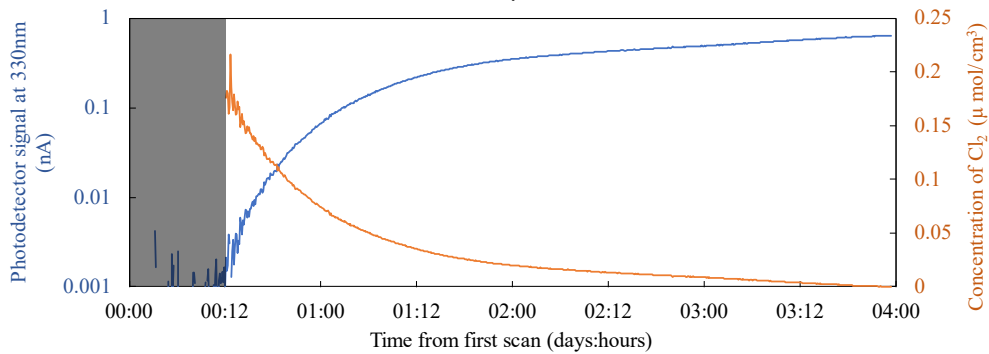

**Fig 3.** The signal measured at 330 nm over time for the fibre under test, and the calculated mean molar concentration in the fibre above the final concentration after 4 days, for the same time period. This trace begins after 12 hrs, as this is when the signal at 330 nm was no longer in the noise floor of our measurements.

The origin of chlorine in hollow cores (in the form of HCl as unambiguously shown by IR spectra [2]) was attributed to the low-OH F300 synthetic fused silica from which the fibre was made, and we make the same assumption here: there were simply no other sources of chlorine in our experiment. The concentration inferred from our measurements is surprising, and we do not propose a mechanism for the evolution of $Cl_2$ from the silica glass. However, the ~1980°C temperature of the glass as the fibre was drawn provides an energetic environment for both chemical processes and diffusion. Previous work on high purity silica glass for fibres suggests that chlorine diffusion out of the glass can be rapid at temperatures above 1100°C [15]. We note that 1450 ppm of chlorine [2] in silica glass corresponds to a volume concentration of Cl atoms of 90 µmol/cm$^3$, far in excess of that needed for the 0.45 µmol/cm$^3$ mean concentration of $Cl_2$ in the hollow core inferred from our measurements, and comparable to the concentration of Cl atoms in pure chlorine gas at NTP. From this inferred concentration at 0 hours, we estimate that the Cl in the core is equivalent to the removal of all Cl from the glass to a depth of 15 nm from all surfaces in contact with the core space. These concentrations are based on the known atomic masses of chlorine, silicon and oxygen.

The quadratic dependence of diffusion time on fibre length and the very high concentration of $Cl_2$ observed in our fibre mean that, for any reasonable length of HCF fabricated from low-OH silica, $Cl_2$ absorption will measurably increase the optical attenuation in this spectral band. Our results in Fig 3 show that even for a very short length of 2 metres, $Cl_2$ affected the fibre transmission for 4 days. Long lengths of HCF are a desirable engineering outcome, but care must be taken when performing transmission measurements as the results will change over time. In our results, although the most dramatic difference is in the UV, the change in transmission is broad, from 190-600 nm. Therefore, even for non-UV guiding fibres, this change over time due to presence of $Cl_2$ will have a significant effect for HCFs at visible wavelengths. It may be possible to avoid this problem by fabricating the fibre from chlorine-free fused silica. However, since chlorine is introduced into fused silica to control absorption at IR wavelengths due to the presence of hydroxyl (OH) [6], use of chlorine-free silica may introduce other problems.

### 4. Conclusion

UV guiding anti-resonant HCF's fabricated from commercially-available low-OH fused silica are shown to exhibit temporary high attenuation in a spectral band centred at ~325nm immediately after fabrication. This spectral feature disappears from the fibre with time. The time-dependent loss is well matched to the absorption spectrum of gaseous $Cl_2$. It appears that the $Cl_2$ sublimates from the glass as the fibre is drawn, and afterwards dissipates by diffusion out of the ends of the fibre and/or reaction with atmospheric water vapour to form HCl. HCl is already known to be present in HCFs fabricated from F300 synthetic fused silica [2]. Even for a short length of fibre (2 metres), it took 4 days for $Cl_2$ absorption to be negligible in the transmission traces. Our findings suggest that hollow-core fibres drawn from low-OH silica with lengths relevant to engineering applications may have optical properties that change over long timescales.

### Funding
This work was funded by the EPSRC under grant EP/T020903/1.

### Disclosures
The authors declare no conflict of interest.

### Data availability
The relevant data is available from [11].

### References

1. F. Yu, M. Cann, A. Brunton, W. J. Wadsworth, and J. C. Knight, "Single-mode solarization-free hollow-core fiber for ultraviolet pulse delivery," Opt. Express, 26(8), 10879–10887, (2018)
2. F. Yu, W. J. Wadsworth, and J. C. Knight, "Low loss silica hollow core fibers for 3–4 μm spectral region," Opt. Express 20, 11153-11158 (2012)
3. I. Gris-Sanchez and J. C. Knight, "Time-Dependent Degradation of Photonic Crystal Fiber Attenuation Around OH Absorption Wavelengths," in Journal of Lightwave Technology 30(23), 3597-3602 (2012)
4. S. Rikimi. Y.Chen, T.W. Kelly, I. A. Davidson, G. T. Jasion, M. Partridge, K. Harrington, T. D. Bradley, A. A. Taranta , F. Poletti, M. N. Petrovich, D. J. Richardson and N. V. Wheeler, "Internal Gas Composition and

Pressure in As-drawn Hollow Core Optical Fibers," in Journal of Lightwave Technology 40(14), 4776-4785 (2022)
5. S. Rikimi, Y. Chen, M. C. Partridge, T. D. Bradley, I. A.K. Davidson, A. A. Taranta, F. Poletti, M. N. Petrovich, David J. Richardson and Natalie V. Wheeler, "Growth of Ammonium Chloride on Cleaved End-Facets of Hollow Core Fibers," in Conference on Lasers and Electro-Optics, OSA Technical Digest (Optica Publishing Group, 2020), paper SF2P.4.
6. L. A. Moore and C. M. Smith, "Fused silica as an optical material [Invited]," Opt. Mater. Express 12, 3043-3059 (2022)
7. R. Mears, K. Harrington, J. C. Knight, W. J. Wadsworth, J. M. Stone, T. A. Birks, "Reversible photodarkening in UV-guiding hollow-core optical fibres," Proc. SPIE PC12882, Optical Components and Materials XXI, PC128820K (9 March 2024)
8. R. Mears K. Harrington, W. J. Wadsworth, J. C. Knight, J. M. Stone and T. A. Birks, "Guidance of ultraviolet light down to 190 nm in a hollow-core optical fibre," Opt. Express 32, 8520-8526 (2024)
9. W. Ding, Y.-Y. Wang, S.-F. Gao, M. Wang and P. Wang, "Recent progress in low-loss hollow-core anti-resonant fibers and their applications," IEEE J. Sel. Top. Quantum Electron. 26(4), 1–12 (2020).
10. G. Jackson, G. T. Jasion, T. D. Bradley, F. Poletti, and I. A. Davidson, "Three stage HCF fabrication technique for high yield, broadband UV-visible fibers," Opt. Express 32, 7720-7730 (2024)
11. K. Harrington, R. Mears, J. M. Stone, W. J. Wadsworth, J. C. Knight, T. A. Birks, "Optical absorption spectrum reveals gaseous chlorine in anti-resonant hollow core fibres," Opt. Express, *in press* (2024)
12. H. Keller-Rudek, G. K. Moortgat, R. Sander and R. Sörensen, "The MPI-Mainz UV/VIS spectral atlas of gaseous molecules of atmospheric interest", Earth Syst. Sci. Data, 5, 365–373, (2013)
13. M. J. Tang, R. A. Cox, and M. Kalberer, "Compilation and evaluation of gas phase diffusion coefficients of reactive trace gases in the atmosphere: volume 1. Inorganic compounds", Atmos. Chem. Phys., 14, 9233–9247, (2014).
14. E. J. Carr, "Characteristic time scales for diffusion processes through layers and across interfaces," Phys. Rev. E, 97, 042115, (2018).
15. J. Kirchhof, S. Unger, B. Knappe, H. -J. Pissler, K. Ruppert and R. Köppler, "Chlorine incorporation into silica lightguide materials," OFC '98. Optical Fiber Communication Conference and Exhibit. Technical Digest. Conference Edition. 1998 OSA Technical Digest Series Vol.2 (IEEE Cat. No.98CH36177), San Jose, CA, USA, 1998, pp. 185-186, doi: 10.1109/OFC.1998.657318.